\documentstyle[epsfig]{aipproc}

\begin{document}
\title{The Galactic Extinction Toward GRB 970228 and Its Implications}
%\\for the Point-Like and Extended Optical Sources}

\author{Francisco J. Castander and D. Q. Lamb}
\address{Department of Astronomy and Astrophysics, University of
Chicago}

\maketitle

%%%%%%%%%%%%%%%%%%%%%%%%%%%%%%%%%%%%%%%%%%%%%%%%%%%%%%%%%%%%%%%% %
\begin{figure}[t]
\centerline{\epsfig{file=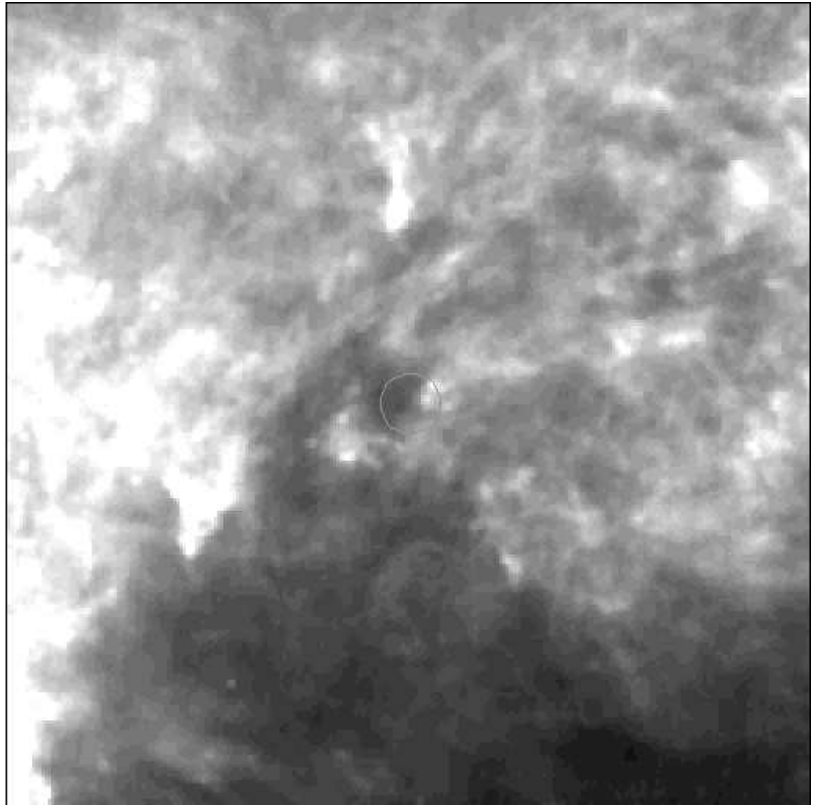,scale=0.59,clip}}
\caption[fooba1]{IRAS 100 $\mu$m map of the GRB 970228 field, covering
$8.5^\circ \times 8.5^\circ$ and having a resolution of $2'$. The bright
regions correspond to strong dust emission, the dark regions to weak
dust emission.  We have superposed on the map a circle $20'$ in radius,
centered on the position of the optical transient (R.A. = $5^{\rm
h}01^{\rm m}46.7^{\rm s}$, Decl. = $+11^\circ46'54''$, J2000).} 
\label{iras_image} \end{figure} %
%%%%%%%%%%%%%%%%%%%%%%%%%%%%%%%%%%%%%%%%%%%%%%%%%%%%%%%%%%%%%%%%

\begin{abstract}
The IRAS 100 micron image of the GRB 970228 field shows that the amount
of galactic dust in this direction is substantial and varies on
arcminute angular scales.  From an analysis of the observed surface
density of galaxies in the $2.6' \times 2.6'$ HST WFPC image of the GRB
970228 field, we find $A_V = 1.1 \pm 0.10$.  From an analysis of the
observed spectra of three stars in the GRB 970228 field, we find $A_V =
1.71^{+0.20}_{-0.40}$. This value may represent the best estimate of
the extinction in the direction of GRB 970228, since these three stars
lie only $2.7''$, $16''$, and $42''$ away from the optical transient.  If
instead we combine the two results, we obtain a conservative value $A_V
= 1.3 \pm 0.2$.  This value is significantly larger than the values
$A_V = 0.4 - 0.8$ used in papers to date.  The value of $A_V$ that we
find implies that, if the extended source in the burst error circle is
extragalactic and therefore lies beyond the dust in our own galaxy, its
optical spectrum is very blue: its observed color $(V-I_c)_{\rm obs}
\approx 0.65^{+0.74}_{-0.94}$ is consistent only with a starburst
galaxy, an irregular galaxy at $z > 1.5$, or a spiral galaxy at $z >
2$.  On the other hand, its observed color and surface brightness
$\mu_V \approx 24.5$ arcsec$^{-2}$ are similar to those expected for
the reflected light from a dust cloud in our own galaxy, if the cloud
lies in front of most of the dust in this direction.
\end{abstract}

%%%%%%%%%%%%%%%%%%%%%%%%%%%%%%%%%%%%%%%%%%%%%%%%%%%%%%%%%%%%%%%%
%
\begin{figure}[t]
\centerline{\epsfig{file=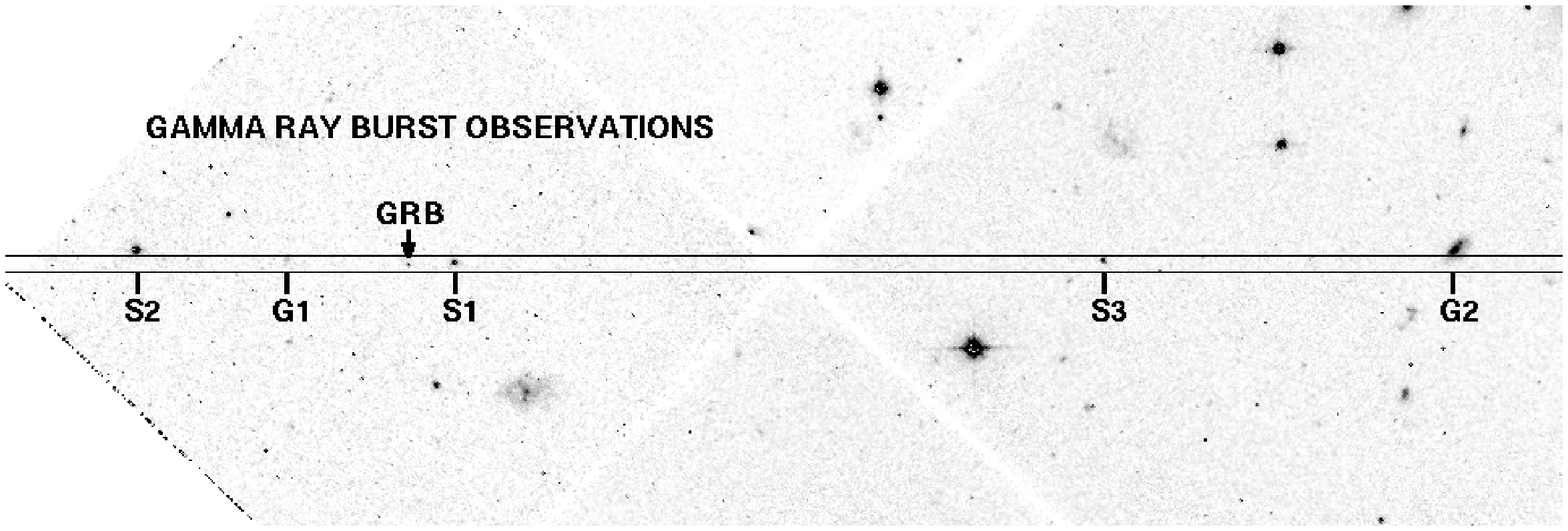,scale=0.57,angle=-90}}
\caption[fooba2]{Position of the $1''$ wide slit used in observations made
on 1997 March 31, April 1, and April 2 UT using the LRIS spectrograph
on the Keck II 10-meter telescope\cite{Tonry97a,Tonry97b}, superposed
on the HST WFPC 606 nm images of the GRB 970228 field obtained using
the HST on 1997 March 26\cite{Sahu97}.  The small truncated square to
the left is the field of the PC chip; the other three truncated squares
are the fields of the WF chips.  The position of the optical transient
is labeled ``GRB;'' also labeled are the positions of stars S1, S2, and
S3; and galaxies G1 and G2.} 
\label{skymap}
\end{figure}
%
%%%%%%%%%%%%%%%%%%%%%%%%%%%%%%%%%%%%%%%%%%%%%%%%%%%%%%%%%%%%%%%%

%%%%%%%%%%%%%%%%%%%%%%%%%%%%%%%%%%%%%%%%%%%%%%%%%%%%%%%%%%%%%%%%
%
\begin{figure}[t]
\centerline{\epsfig{file=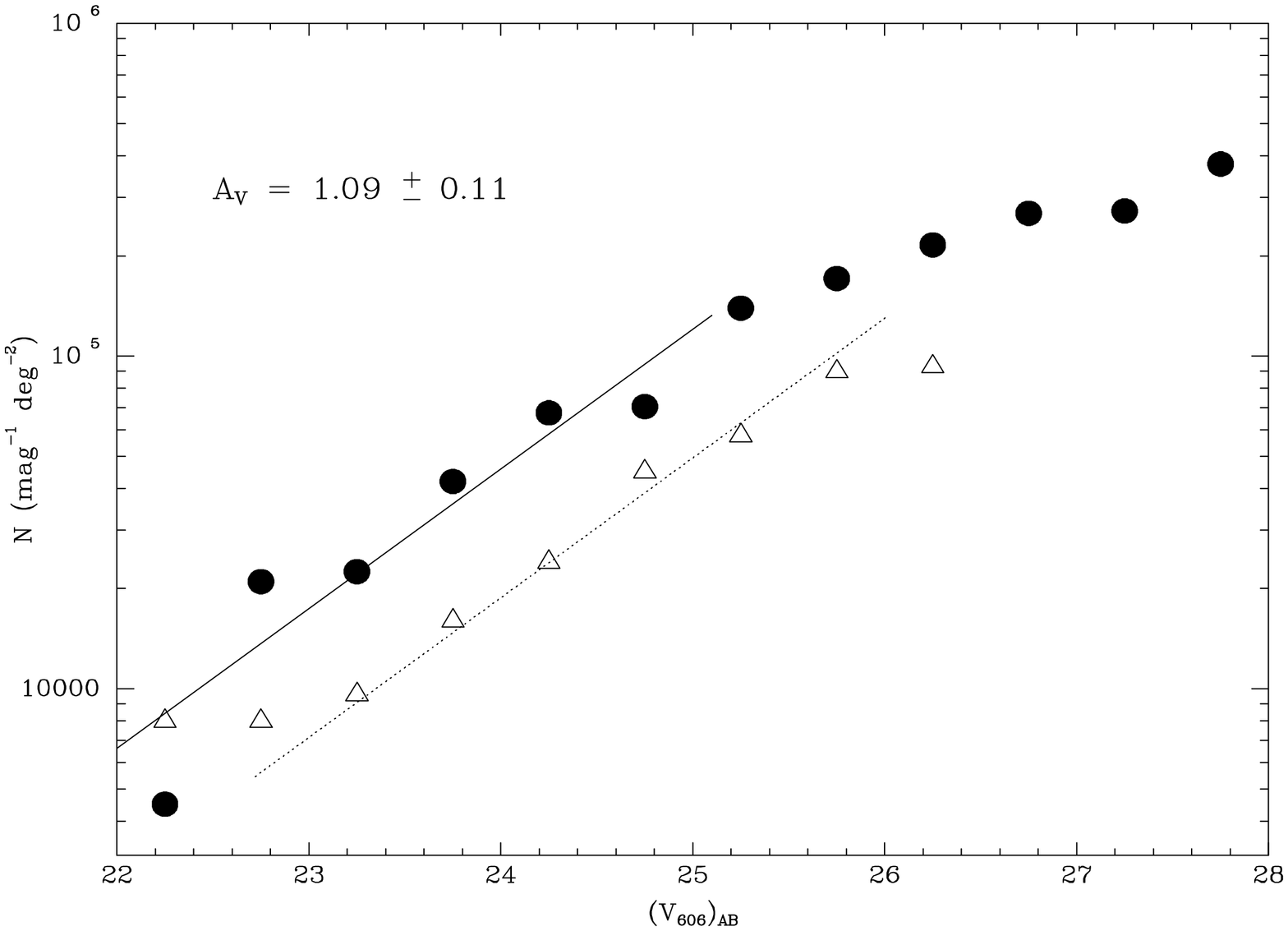,scale=0.32,angle=90}
            \epsfig{file=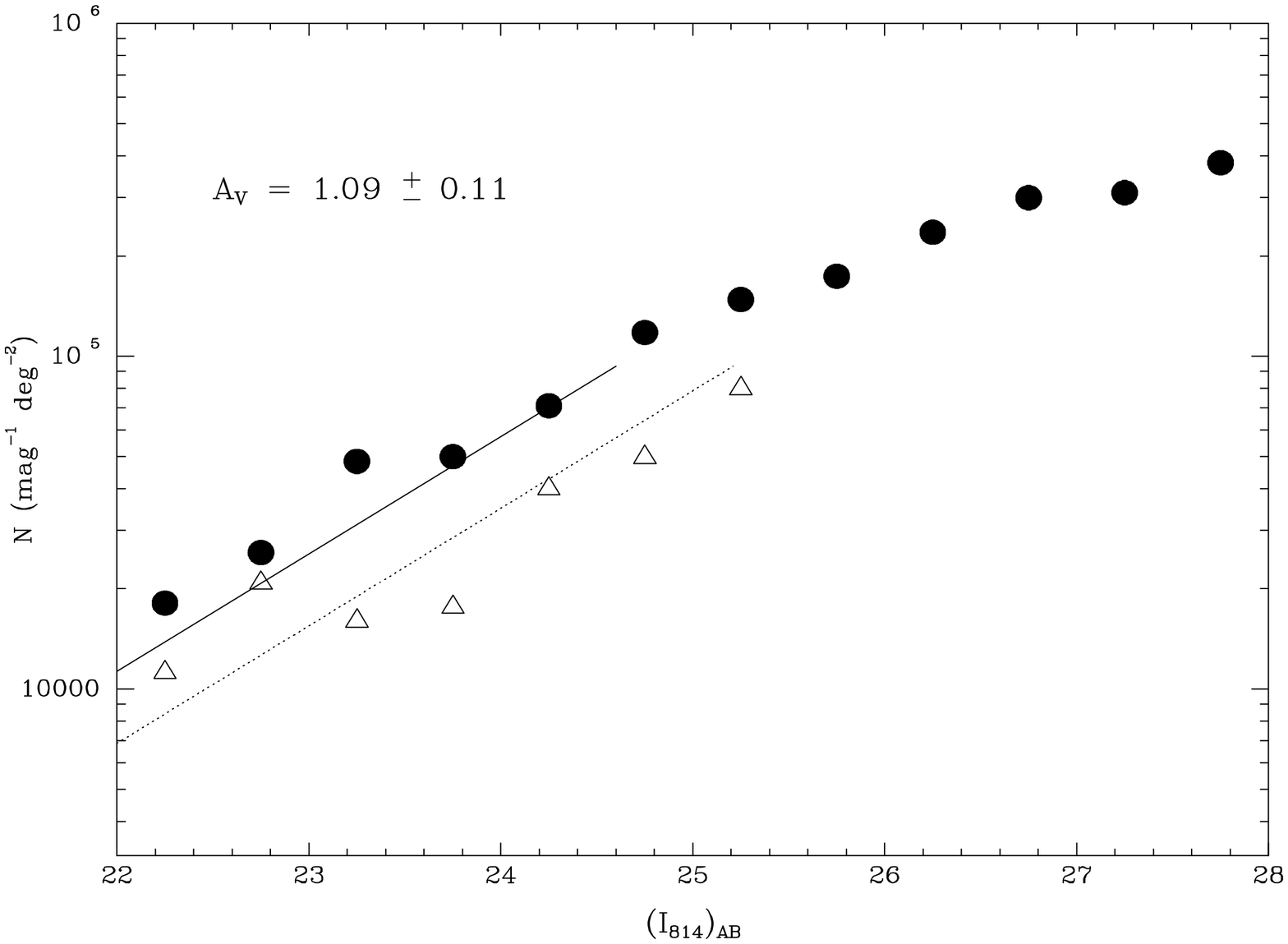,scale=0.32,angle=90}}
\centerline{\epsfig{file=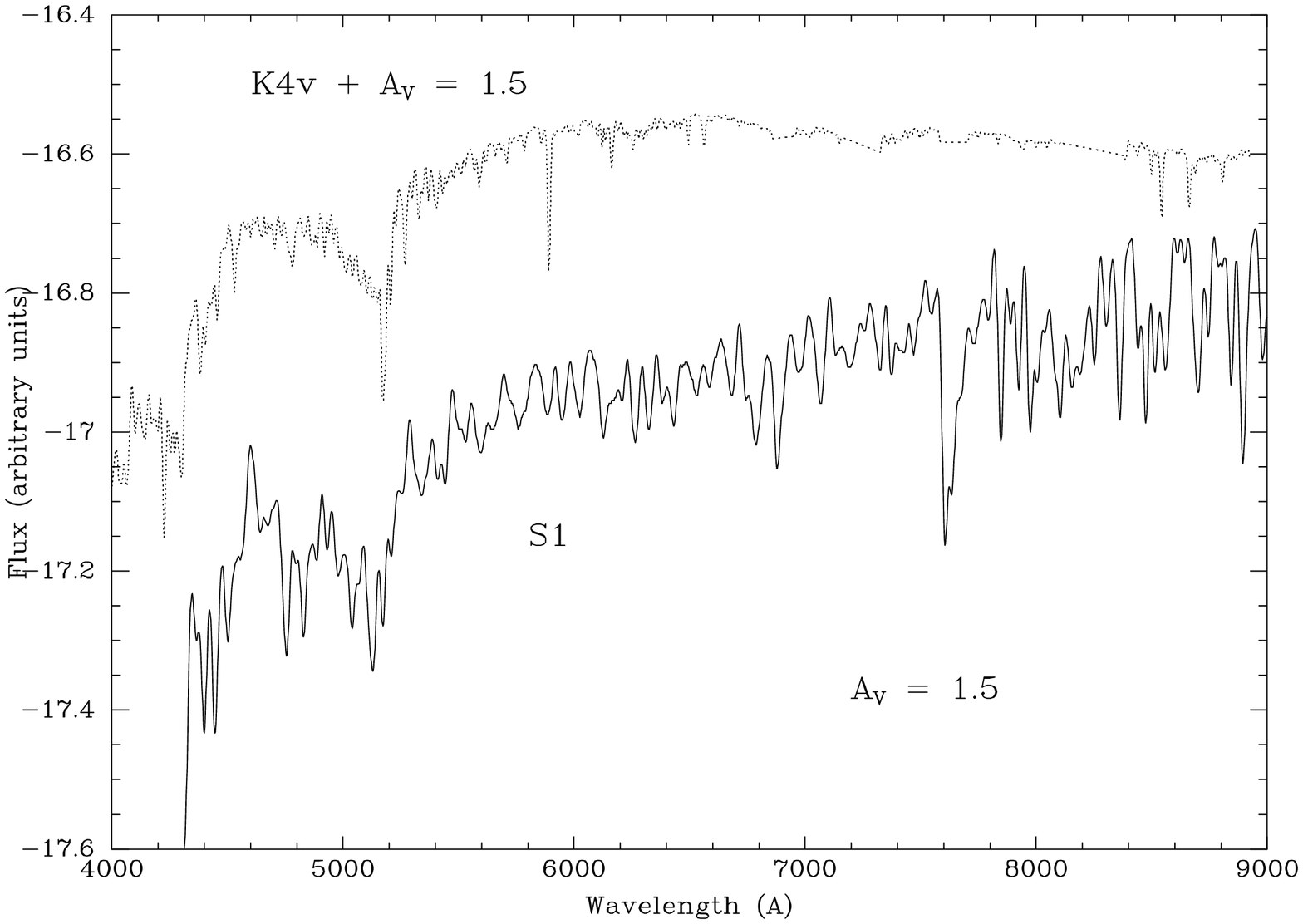,scale=0.32,angle=90}
            \epsfig{file=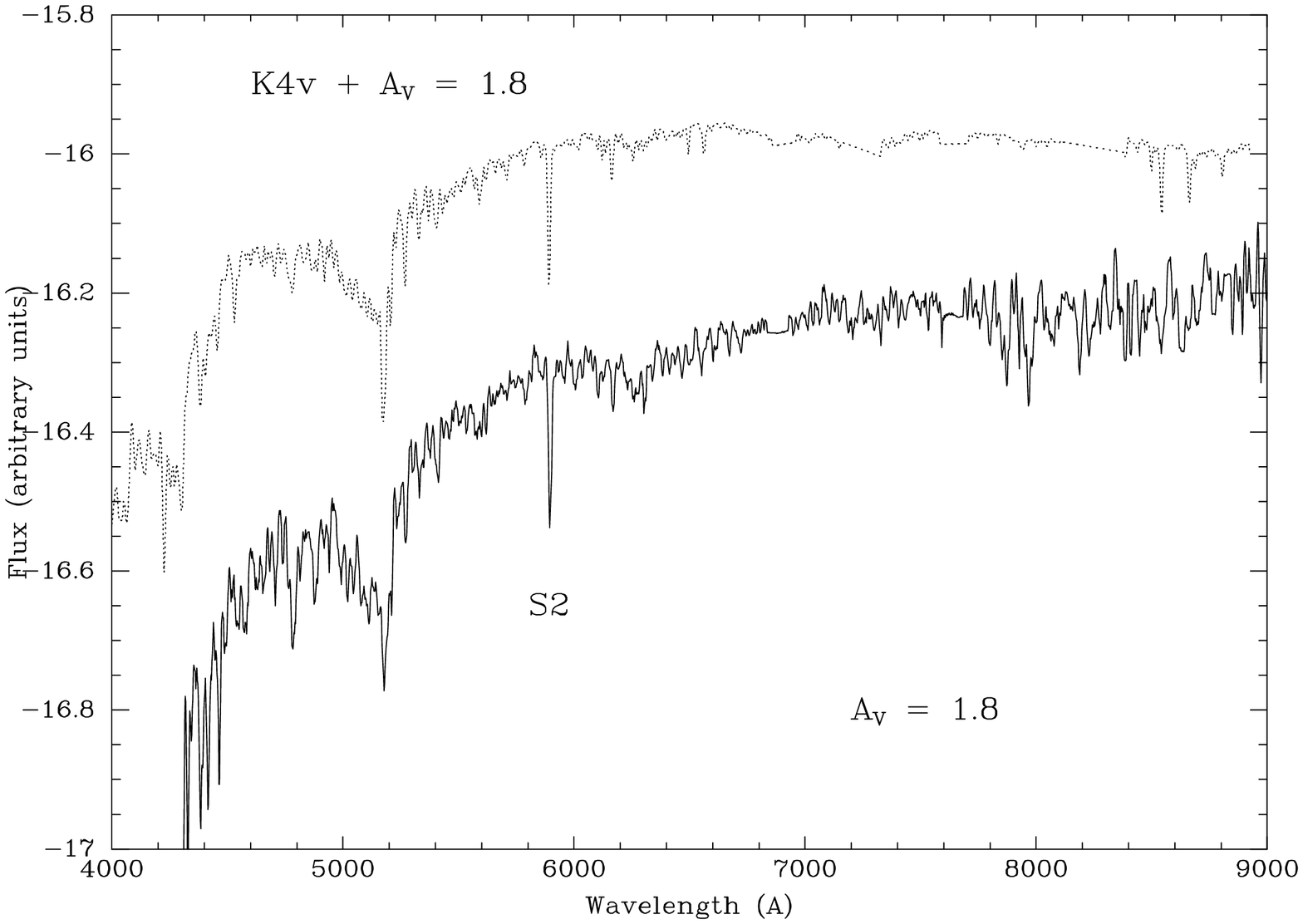,scale=0.32,angle=90}}
\caption[fooba]{Upper panels:  Surface density of galaxies as a
function of apparent magnitude in the WFPC 606 nm (left panel) and 814
nm (right panel) images of the HDF (filled circles) and the GRB 970228
field (open triangles).  Also shown is the surface density of galaxies
as a function of apparent magnitude for the HDF (solid lines) and the
GRB 970228 field (dotted lines), given by the  best-fit model we use to
describe the combined data for the HDF and GRB 970228 fields.

Lower panels:  Observed (solid curves) and model (dotted curves)
spectra of stars S1 (left panel) and S2 (right panel).  The observed
spectra of S1 and S2 are best fit by that of a K4v star, reddened by
$A_V = 1.4^{+0.5}_{-0.8}$ and $A_V = 1.8^{+0.2}_{-0.5}$, respectively.}
\label{skyplot}
\end{figure}
%
%%%%%%%%%%%%%%%%%%%%%%%%%%%%%%%%%%%%%%%%%%%%%%%%%%%%%%%%%%%%%%%%
%\pagebreak

\section*{Introduction}

The IRAS 100 micron image of the GRB 970228 field shows that the amount
of galactic dust in this direction is substantial and varies
significantly on arcminute angular scales (see Figure 1).  Here we
report a determination of the visual extinction $A_V$ toward GRB 970228
using two methods: (1) the observed versus the expected surface density
of galaxies in the HST WFPC 606 and 814 nm images of the GRB 970228
field; and (2) the observed color versus the spectral type of three
stars that lie near the position of the optical transient (see Figure
2). 
\section*{Analysis and Results}

Galaxy number counts can be used to measure directly the relative
extinction between two fields.  Since absorption due to dust increases
the observed apparent magnitude of a galaxy at infrared, optical, and
ultraviolet wavelengths, the number of galaxies per unit area brighter
than a given apparent magnitude is reduced if extinction is present. We
have compared the surface density of galaxies as a function of apparent
magnitude in the HST WFPC 606 and 814 nm images of the GRB 970228
field\cite{Sahu97} with similar images of the Hubble Deep
Field\cite{Williams96} (HDF) (Figure 3). We have used the same
procedure to analyze both fields; our results for the HDF field agree
with those reported earlier\cite{Williams96}.  We restrict our
comparison to galaxies in the WFPC 606 and 814 nm images of the GRB
970228 field for which $V_{606} \le 26.0$ and $I_{814} \le 25.2$,
apparent magnitudes for which the galaxy counts in this field are
complete.

We determine the extinction in the HST WFPC images of the GRB 970228
field using a maximum likelihood method.  We construct a likelihood
function that is a product of four individual likelihood functions, one
for the 606 and 814 nm images of each field.  We model the surface
density of galaxies as a function of apparent magnitude in each image
as a power-law distribution.  We assume that the slopes $\alpha_{606}$
and $\alpha_{814}$ of the power-law distributions in the 606 and 814 nm
images are the same in the GRB 970228 field and HDF.  We further assume
that the extinction in both fields is described by the typical
extinction behavior of the interstellar medium, so that $A_{606} = 0.92
A_V$ and $A_{814} = 0.61 A_V$.  The resulting likelihood function only
depends on five parameters:  the normalizations and the slopes
$\alpha_{606}$ and $\alpha_{814}$ of the power-law distributions in the
two filters and the difference in the extinction between the GRB 970228
field and the HDF.  We then marginalize this likelihood function over
the normalizations in the 606 nm and 814 nm filters and use the
measured value $A_V^{HDF} = 0.0$\cite{Williams96}.  This reduces the
number of parameters to three: $\alpha_{606}$, $\alpha_{814}$, and
$A_V$. Maximizing the likelihood of the observed surface density of
galaxies as a function of apparent magnitude in each image of each
field, given the model, we obtain a best-fit value $A_V = 1.1 \pm
0.10$. 

%%%%%%%%%%%%%%%%%%%%%%%%%%%%%%%%%%%%%%%%%%%%%%%%%%%%%%%%%%%%%%%%
%
\begin{figure}
\centerline{\epsfig{file=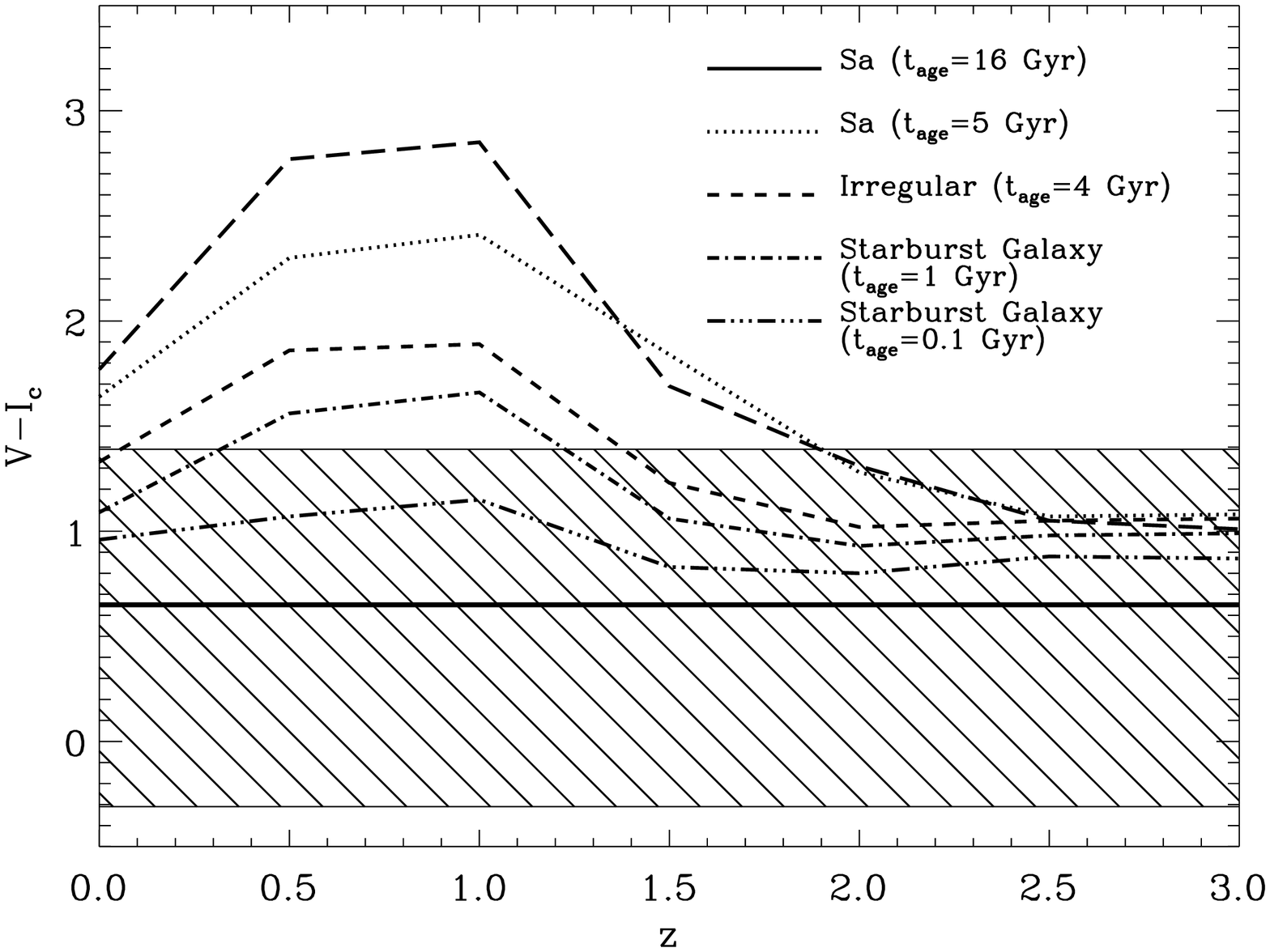,scale=0.4,angle=90}}
\caption[fooba]{Expected colors for several kinds of galaxies at
different ages as a function of redshift $z$.  The curves include the
appropriate K-correction, but assume no galaxy evolution.  The thick
solid line shows $(V-I_c)_{\rm obs}$ for the extended source in the
burst error circle; the hatched region indicates the range of
uncertainty in this color.  This figure shows that the extended source
is very blue: its $V-I_c$ color is consistent only with that expected
for starburst galaxies, for irregular galaxies at $z > 1.5$, or for
spiral galaxies at $z > 2$.}
\label{galaxies}
\end{figure}
%
%%%%%%%%%%%%%%%%%%%%%%%%%%%%%%%%%%%%%%%%%%%%%%%%%%%%%%%%%%%%%%%%

We also determine the visual extinction $A_V$ toward GRB 970228 using
the spectra of three stars, denoted S1, S2, and S3, that lie near the
optical transient in the GRB 970228 error circle (see Figure 2).  The
spectra of these stars were obtained on 1997 March 31, April 1, and
April 2 UT using the LRIS spectrograph on the Keck II 10-meter
telescope\cite{Tonry97a,Tonry97b}.  Comparing the observed spectra
with the stellar spectral atlases\cite{Jacoby84,Silva92}, we find that
S1, S2, and S3 lie in the spectral ranges K3v-K5v, K4v-K7v, and
M0v-M3v, respectively.  Reddening the spectra in the stellar spectral
atlases, we obtain best-fit spectra and visual extinction values for
S1, S2, and S3 of K4v and $A_V = 1.4^{+0.5}_{-0.8}$, K4v and $A_V =
1.8^{+0.2}_{-0.5}$, and M2v and $A_V = 1.8^{+0.5}_{-1.0}$,
respectively.  Weighting the individual values of $A_V$ by the
signal-to-noise of their spectra, we obtain a combined value $A_V =
1.71^{+0.20}_{-0.40}$.  This may represent the best estimate of the
extinction in the direction of GRB 970228, since these three stars lie
only $2.7''$, $16''$, and $42''$ away from the position of the optical
transient.

Combining the results of our analysis of the surface
density of galaxies in the GRB 970228 field and the results of our
analysis of the spectra of stars S1, S2, and S3, we obtain a
conservative value $A_V = 1.3 \pm 0.2$.
This value is consistent with the X-ray spectrum of the gamma-ray
burst  itself, which yields $n_H = 3.5^{+3.3}_{-2.3} \times 10^{21}$
cm$^{-2}$\cite{Costa97}, implying $A_V = 2.3^{+2.1}_{-1.5}$.

\section*{Conclusions}

From an analysis of the observed surface density of galaxies in the
$2.6' \times 2.6'$ HST WFPC image of the GRB 970228 field, we find $A_V
= 1.1 \pm 0.10$.  From an analysis of the observed spectra of three
stars in the GRB 970228 field, we find $A_V = 1.71^{+0.20}_{-0.40}$.
This value may represent the best estimate of the extinction in the
direction of GRB 970228, since these three stars lie only $2.7''$,
$16''$, and $42''$ away from the optical transient.  If instead we
combine the two results, we obtain a conservative value $A_V = 1.3 \pm
0.2$.

This value is significantly larger than the values $A_V = 0.4 - 0.8$
used in papers to date.  The value of $A_V$ that we find implies that,
if the extended source in the burst error circle is extragalactic, its
optical spectrum is very blue: its observed color of $(V-I_c)_{\rm obs}
\approx 0.65^{+0.74}_{-0.94}$ is consistent only with a starburst
galaxy, an irregular galaxy at $z > 1.5$, or a spiral galaxy at $z >
2$, after taking into account the measured extinction.  On the other
hand, its observed color and surface brightness $\mu_V \approx 24.5$
arcsec$^{-2}$ are similar to those expected for the reflected light
from a dust cloud in our own galaxy, if the cloud lies in front of most
of the dust in this direction.
\medskip

We thank John Tonry, Esther Hu, Len Cowie, and Richard McMahon for
making available the spectra 
%that they obtained 
of the galaxies
and stars in the GRB 970228 field.  We acknowledge support from NASA
grants NAGW-4690 and NAG 5-1454.
%, and NAG 5-4406.


\begin{references}
%
\bibitem{Sahu97} Sahu, K.~C. et  al. 1997, Nature, 387, 476
%
\bibitem{Williams96} Williams, R.~E. et al. 1996, AJ, 112, 1335
%
\bibitem{Tonry97a} Tonry, J.~L., Hu, E.~M., Cowie, L.~L. and McMahon, R.~G. 
1997a, IAU Circular No. 6620
%
\bibitem{Tonry97b} Tonry, J.~L. et al.
%, Hu, E.~M., Cowie, L.~L. and McMahon, R.~G. 
1997b, http://www.ifa.hawaii/faculty/hu/grb.html
%
\bibitem{Jacoby84} Jacoby, G.~H., Hunter, D.~A. \& Christian, C.~A.
1984, ApJS, 56, 257
%
\bibitem{Silva92} Silva, D.~R. \& Cornell, M.~E. 1992, ApJS, 81, 865
%
\bibitem{Costa97} Costa, E. et al. 1997, Nature, 387, 783
\end{references}
\end{document}